\documentclass[preprint,authoryear,12pt]{elsarticle}
\pdfoutput=1  

\usepackage[T1]{fontenc}
\usepackage[utf8]{inputenc}
\usepackage[english]{babel}

\usepackage[pdftex]{color}
\usepackage[pdftex,colorlinks=true]{hyperref}

\usepackage{amsfonts}         
\usepackage{amsmath}
\usepackage{amssymb}
\usepackage{amsthm}

\usepackage{subfig}

\newcommand{\Z}{\mathbb{Z}}
\newcommand{\TODO}[1]{}
\newcommand{\A}{\mathrm{A}}
\newcommand{\B}{\mathrm{B}}
\newcommand{\xa}{\mathrm{a}}
\newcommand{\xb}{\mathrm{b}}
\newcommand{\xg}{\mathrm{g}}
\newcommand{\iverson}[1]{\mathbf{1}_{\{#1\}}}

\journal{Journal of Theoretical Biology}

\begin{document}

\begin{frontmatter}

\title{Stable trimorphic coexistence in a lattice model of spatial competition with two site types}

\date{October 13, 2011}

\author{Ilmari Karonen\corref{ikaronen}}
\ead{ilmari.karonen@helsinki.fi}
\address{Department of Mathematics and Statistics, University of Helsinki, Finland}

\cortext[ikaronen]{\textit{Address for correspondence:} Department of Mathematics and Statistics,
  University of Helsinki, PO Box 68, FI-00014, Finland. Phone: +358-41-456 3263, fax: +358-9-1951
  1400.}

\begin{abstract}
I examine the effect of exogenous spatial heterogeneity on the coexistence of competing species
using a simple model of non-hierarchical competition for site occupancy on a lattice.  The sites on
the lattice are divided into two types representing two different habitats or spatial resources.
The model features no temporal variability, hierarchical competition, type-dependent interactions
or other features traditionally known to support more competing species than there are resources.
Nonetheless, stable coexistence of two habitat specialists and a generalist is observed in this
model for a range of parameter values.  In the spatially implicit mean field approximation of the
model, such coexistence is shown to be impossible, demonstrating that it indeed arises from the
explicit spatial structure.
\end{abstract}


\end{frontmatter}

\newpage

\section{Introduction}

The coexistence of competing species, and factors promoting and limiting it, are of considerable
practical and theoretical interest in ecology.  A well known ``rule of thumb'' is the principle of
competitive exclusion \citep{gause1932, tilman1982, levin1970}, which states that at most $n$
mutually competing species may stably coexist on $n$ available resources.

Under suitable assumptions, the competitive exclusion principle can be proven as a mathematical
theorem.  However, if these assumptions are violated --- for example, if the resource abundances may
fluctuate over time, either due to external resources or simply because the ecological dynamics tend
to a cyclic or chaotic attractor --- it may no longer hold (although a related concept, the
essential dimensionality of the environment \citep{dieckmann2006, metz2008}, may still be applied to
such systems).

For systems which do satisfy these assumptions, the validity of the competitive exclusion principle
depends fundamentally on just what we count as ``a resource'' \citep{abrams-88}.  This is not as
simple a matter as it sounds.  Were one to ask a practical-minded ecologist what constitutes a
resource, they might name examples such as water, sunlight and nutrients for plants, or prey species
for animals.  But in the mathematically exact form of the competitive exclusion principle, almost
anything may constitute a distinct resource: a single prey individual, a square meter of land, a
specific combination of nutrient concentrations, etc.\ Thus, even in systems which the competitive
exclusion principle formally holds, the actual number of potentially coexisting competitors may be
greater than one would expect by naively undercounting the resources.

The effect of spatial structure on the maximum diversity a system can support is, in particular,
often neglected.  For example, systems consisting of several distinct types of habitats are often
modelled by assuming that each habitat constitutes a homogeneous patch, within which the populations
are well mixed.  If competition between individuals is for suitable living space within these
habitats, space in each habitat then becomes a single resource, and thus one would expect (and will,
given these assumptions, mathematically discover) that at most $n$ competitors may stably persist in
$n$ distinct habitats.

In reality, however, even if habitats are homogeneous, they are certainly not usually well mixed.
Thus, individuals living near habitat boundaries will, over time, experience a different environment
than those living in the interior of habitat patches.  (Even if the individuals themselves don't
move or interact with anything outside their local site, their offspring must still disperse and may
end up in a different habitat.)  This can create additional niches near habitat boundaries in which
additional competing species might be able to coexist.

To demonstrate this, I present in this paper a simple spatially explicit toy model of site occupancy
competition, which supports stable coexistence of three strains -- two specialists and one
generalist -- on two spatially segregated habitats on a lattice of sites.  This model contains no
other features known to promote coexistence, such as internally or externally generated temporal
fluctuation \citep{hsu1978, armstrong1980}, hierarchical site competition \citep{adler2000,
  tilman1994}, direct strain-dependent competition terms \citep{murrell-law-03} or cooperative or
other nonlinear interactions between individuals.  Rather, the coexistence arises merely from the
presence of habitat boundaries combined with passive distance-limited dispersal, which causes
specialist strains to be locally maladapted near these boundaries and thereby allows more generalist
strains to persist there.

As far as I know, this particular mechanism of coexistence has not been previously studied in
individual-based models.  A similar coexistence mechanism was very recently described in a
reaction--diffusion model and in a 1D stepping stone model by \citet{debarre2011}, who termed it
``habitat boundary polymorphism''.  The results in this paper parallel theirs, and confirm that this
mechanism is robust with respect to the details of the model, provided that the basic features of
habitat heterogeneity and passive distance-limited dispersal are present.

A model almost identical to mine was studied by \citet{lanchier2006}, who showed analytically that
it could support stable dimorphic coexistence, either of a generalist and a specialist strain or of
two different specialist strains.  My model differs from theirs only in that they restrict the
habitat configuration to the special case of an infinite regular checkerboard pattern consisting of
$n$-by-$n$ squares of each habitat type.\footnote{\Citet{lanchier2006} also consider a somewhat
  different set of dispersal kernels, but both theirs and mine include the basic case of strict
  nearest-neighbor dispersal.} However, while Lanchier and Neuhauser did briefly remark that ``the
generalists persist for a very long time along the boundaries \ldots where the density of
specialists is low'' in numerical simulations, they do not seem to have investigated this
possibility of trimorphic coexistence in their model further.  Similarly, \citet{snyder-chesson-03}
define a model quite similar to mine (although in discrete time and one spatial dimension), and
observe the enhancing effect of stable spatial heterogeneity and local dispersal on coexistence of
competitors, but also restrict their analysis to two competitors.

\section{Model definition}
I model a population of haploid, asexually reproducing sessile organisms with distance-limited
offspring dispersal.  The model I'll define below belongs to the class of lattice contact processes
\citep{harris1974, neuhauser1992}, in which the environment is taken to consist of a lattice of
discrete sites, and interactions are (mainly) between nearest neighbor sites.

Let $L$ be a regular two-dimensional lattice of sites (e.g. $L = \Z^2$, although for numerical
simulations a finite lattice must obviously be used).  To each site I assign a random, fixed habitat
class (``A'' or ``B''), such that both classes of sites are present in $L$ in equal numbers.  (I
will describe the way in which the habitat classes are assigned in more detail below.)

Each site in $L$ may, at a given time, be either vacant or occupied by an individual belonging to
one of three strains: an A-specialist ($a$), a B-specialist ($b$) or a generalist ($g$).  The two
specialist strains can only occupy sites in their respective habitat class, while the generalist
strain may occupy any site.

Except for their different habitat adaptation, the strains are completely identical: all individuals
die with rate $\mu$ and produce offspring with rate $\phi$.  Offspring are randomly dispersed to the
eight nearest sites surrounding the parent individual's site (or possibly, with probability
$\epsilon$, to a randomly chosen site in $L$), and will become new individuals if the site they land
in is vacant and of a suitable habitat class.  However, the generalists pay a cost for their ability
to live in either habitat: their offspring survive only with probability $p_{\xg} <
1$.\footnote{Equivalently, I could've scaled the offspring production rate of the generalists to
  $p_{\xg} \phi$.  However, the definition I've chosen allows a straightforward generalization to
  semi-specialist strains with different (non-zero) survival rates in different habitats.}

The time evolution of the entire lattice $L$ can thus be considered as a continuous-time Markov
process, whose state at time $t$ is a function $\eta_t : L \to \{0, \xa, \xb, \xg\}$ mapping sites
in $L$ to their occupancy state (with the state $0$ denoting a vacant site).  The local transition
rates of a site $x$ are then
\begin{equation}
\begin{array}{r @{\hspace{4pt}=\hspace{4pt}} r}
  r(0 \to \xa) & \displaystyle \iverson{H_x=\A} \phi \left(
    (1-\epsilon) \sum_{y \in E_x} \frac{\iverson{\eta_t(y)=\xa}}{|E_y|} +
       \epsilon  \sum_{y \in L}  \frac{\iverson{\eta_t(y)=\xa}}{|L|}
  \right), \\
  r(0 \to \xb) & \displaystyle \iverson{H_x=\B} \phi \left(
    (1-\epsilon) \sum_{y \in E_x} \frac{\iverson{\eta_t(y)=\xb}}{|E_y|} +
       \epsilon  \sum_{y \in L}  \frac{\iverson{\eta_t(y)=\xb}}{|L|}
  \right), \\
  r(0 \to \xg) & \displaystyle p_{\xg} \phi \left(
    (1-\epsilon) \sum_{y \in E_x} \frac{\iverson{\eta_t(y)=\xg}}{|E_y|} +
       \epsilon  \sum_{y \in L}  \frac{\iverson{\eta_t(y)=\xg}}{|L|}
  \right),
\end{array}\label{eqn-transitions}
\end{equation}
and
\[
  r(\xa \to 0) = r(\xb \to 0) = r(\xg \to 0) = \mu,
\]
where $r(s \to s')$ is the rate at which a site in state $s$ changes to state $s'$, $H_x$ is the
habitat class of site $x$ and $E_x$ is the set of sites adjacent to $x$.

The behavior of the model is determined by the habitat configuration along with two parameters: the
scaled baseline fecundity rate $\lambda = \phi/\mu$ (or equivalently, the scaled mortality rate
$\lambda^{-1} = \mu/\phi$) and the generalist survival rate $p_{\xg}$.  The baseline fecundity
affects the equilibrium population density of both the specialists and the generalist strain: at low
$\lambda$ all strains die out, while at very high $\lambda$ almost all sites are occupied at any
time.

The parameter $0 < p_{\xg} < 1$ determines the penalty which the generalist must pay for its ability
to exploit both site types, and is (together with the habitat configuration) crucial in determining
the outcome of the model.  If $p_{\xg}$ is too low, the generalist strain will not be viable, or,
even if viable on its own, may lose in competition to the specialists.  Conversely, if $p_{\xg}$ is
close to 1, the specialist strains gain little or no advantage over the generalist from their
specialization, while paying a considerable price in being able to live in only one habitat, and can
thus be expected to lose to the generalist.

\section{Landscape generation}\label{landscape}
An issue so far overlooked in the model definition above is the way in which the lattice sites are
assigned to their habitat classes.  The simplest way to do so, of course, is to simply assign each
site independently to either habitat with equal probability.  This produces a lattice with no
correlations between the habitat classes of different sites.

However, real environmental features are often correlated, and it would be desirable to consider the
effects of such correlations on the behavior of the model.  To first order, such correlations can be
characterized by the pairwise correlation probability $k = \mathrm{Pr}[ H_x=H_y | y \in E_x ]$,
i.e. the probability that two randomly chosen adjacent sites have the same habitat type.  With an
equal number of sites in each habitat, the pairwise correlation probability on a completely random,
uncorrelated lattice is $k = \frac{1}{2}$, while lattices with $\frac{1}{2} < k < 1$ are positively
correlated and those with $0 < k < \frac{1}{2}$ are anticorrelated.\footnote{On a square lattice
  where each site is adjacent to its 8 nearest neighbors, the smallest achievable value of $k$ is
  $\frac{1}{4}$.}

To generate random habitat class distributions with a given pairwise correlation probability for
numerical simulations, I start by randomly choosing half of the sites and assigning them to habitat
A and the rest to habitat B.  I then apply an iterative annealing process, which reassigns randomly
chosen sites to new habitat classes with suitable weighted probabilities, until the desired value of
$k$ is reached.

There are many possible variations of the general annealing scheme I've used.  The basic idea in all
of them is to change the habitat classes of randomly chosen sites if this would change $k$ in the
desired direction, while occasionally also allowing changes in the other direction so that the
process doesn't get stuck at a local minimum or maximum.

The particular annealing algorithm I've used to generate habitat configurations for the simulations
in this paper\footnote{The interactive Java applets from which the snapshots in figure \ref{fig4sim}
  are taken use a different annealing algorithm.} chooses random adjacent pairs of sites, and swaps
their habitat classes with probability \[ p = \frac{d^\gamma}{d^\gamma + (1-d)^\gamma} \] if
$k_{\mathrm{current}} < k_{\mathrm{target}}$, or with probability $1-p$ otherwise, where $d$ is the
fraction of sites adjacent to the chosen pair which belong to the opposite habitat than their
neighbor in the chosen pair.

The exponent $\gamma$ is a free parameter which controls the ``temperature'' of the process.  When
$\gamma = 1$, the probability of exchanging the habitat classes of a chosen site pair is a linear
function of $d$.  This tends to produce fairly slow convergence and rough, jagged cluster
boundaries.  At high values of $\gamma$, $p$ approaches a step function, producing faster initial
convergence and smoother cluster boundaries, but also increasing the risk of the process getting
stuck at a local maximum or minimum of $k$.

\section{Mean field approximation}
Classical ecological theory predicts that the only possible outcome of three distinct strains
competing for the occupancy of two habitats should be the eventual extinction of one or more of the
strains, except at specific degenerate choices of the model parameters where neutral coexistence may
occur.  I will demonstrate below that this prediction indeed holds if the populations are assumed to
be well mixed, either globally or within each habitat.  However, I shall also show that, in the full
model with explicit spatial structure, a region of stable trimorphic coexistence does exist for
intermediate values of $p_{\xg}$.

Assuming that all offspring disperse uniformly over the entire lattice, i.e. that $\epsilon = 1$ in
\ref{eqn-transitions}, the transition rates of each site are fully described by the mean population
densities $n_\xa$, $n_\xb$ and $n_\xg$ of the different strains, where
\[ n_s = \sum_{y \in L} \frac{\iverson{\eta(y) = s}}{|L|} \]
for each strain $s$.  Further letting the lattice size $|L|$ tend to infinity, one obtains a simple
system of ordinary differential equations describing the time evolution of these mean population
densities---the so called \textit{mean field} approximation---which may be solved analytically.
Such an approximation of an essentially equivalent model was presented by \citet{lanchier2006}, who
showed that trimorphic coexistence was only possible for degenerate choices of parameter values.

However, simply assuming all dispersal to be global completely neglects not only the detailed
spatial habitat structure, but even the pairwise correlation parameter $k$.  A better approximation,
similar to the ``improved mean field approximation'' of \citet{hiebeler-morin-07}, is obtained by
assuming well mixing only within each habitat.  The resulting approximation can be interpreted as a
model of a population inhabiting two well-mixed habitat patches, with a fraction $k$ of all
offspring remaining within their parent's patch and the rest dispersing to the other patch.  This
\textit{two-patch approximation} takes into account the habitat correlation parameter $k$ but still
retains the analytical tractability of the mean field approximation.  (For $k=\frac{1}{2}$, the two
approximations are exactly equivalent.)  Below, I will analyze this approximation of the model
defined above, and show that it also only supports non-degenerate coexistence of at most two
strains.

\begin{figure}
  \centering
  \includegraphics[width=0.4\textwidth]{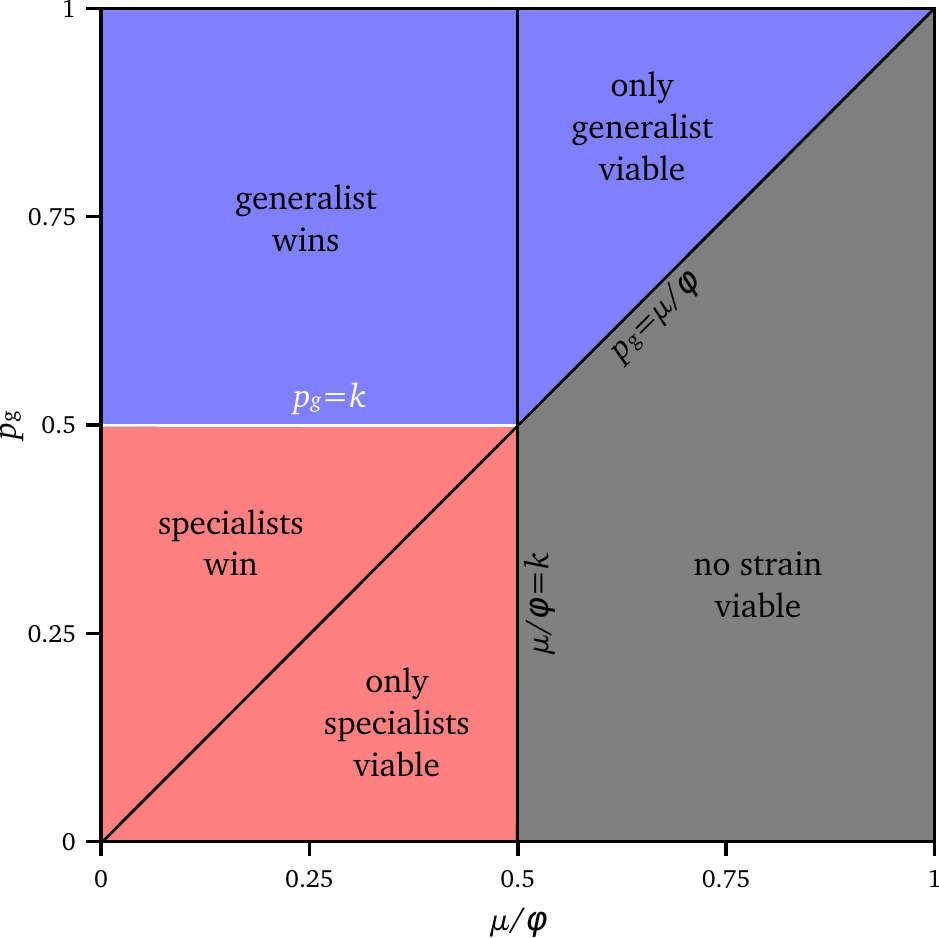}
  \caption{The outcome of the model as predicted by the two-patch approximation for different values
    of the normalized mortality rate $\mu/\phi$ and the generalist survival probability $p_{\xg}$.
    The diagonal line at $\mu/\phi = p_{\xg}$ and the vertical line at $\mu/\phi = k$ mark the
    boundaries at which the generalist and specialist strains go extinct even in the absence of
    competitors.  In the regions marked ``generalist wins'' and ``specialists win'', all strains can
    survive in the absence of competitors, but from any initial state with all three strains
    present, the system converges to either a monomorphic generalist-only state or a dimorphic
    specialist-only state.  Along the white line at $p_{\xg} = k$, the three strains can coexist
    neutrally.}
  \label{fig-meanfield}
\end{figure}

Assume that the occupancy states of the lattice sites are independent, and that the probability of a
site being occupied by a given strain $s$ is equal to the mean population density $n_{s,H}$ of that
type in the site's habitat $H$.  Then, in the limit as $|L| \to \infty$, the time evolution of the
population densities can be written as
\begin{align*}
  \frac{d}{dt} n_{s,\A} &= p_{s,\A} v_{\A} \phi (kn_{s,\A} + (1-k)n_{s,\B}) - \mu n_{s,\A}, \\
  \frac{d}{dt} n_{s,\B} &= p_{s,\B} v_{\B} \phi (kn_{s,\B} + (1-k)n_{s,\A}) - \mu n_{s,\B},
\end{align*}
for $s \in \{\xa, \xb, \xg\}$, where $v_\A$ and $v_\B$ are the vacant site densities in the two
habitats and $p_{s,\A}$ and $p_{s,\B}$ are the probabilities of an offspring of type $s$ surviving
in the respective habitats:
\begin{align*}
  p_{\xa,\A} = p_{\xb,\B} &= 1, \\
  p_{\xb,\A} = p_{\xa,\B} &= 0, \\
  p_{\xg,\A} = p_{\xg,\B} &= p_\xg.
\end{align*}
Equivalently, this system may be written in matrix form as \[ \frac{d}{dt} \bar n_s = M_s \bar
n_s \] for $s \in \{\xa, \xb, \xg\}$, where $\bar n_s = [n_{s,\A}, n_{s,\B}]^T$ and
\[
  M_s = \left[ \begin{array}{cc}
    \phi k p_{s,\A} v_{\A} - \mu & \phi (1-k) p_{s,\A} v_{\A}   \\
    \phi (1-k) p_{s,\B} v_{\B}   & \phi k p_{s,\B} v_{\B} - \mu
  \end{array} \right].
\]
If this system has a nontrivial interior equilibrium $\tilde n_s$, this necessarily implies that
$M_s \tilde n_s = [0, 0]^T \ne \tilde n_s$, and therefore that $M_s$ must be singular, and thus have
a zero determinant, for each strain $s$ present in the population.  Writing out the determinant as
\begin{equation*}
  \begin{aligned}
    |M_s| &= \phi^2 (2k-1) p_{s,\A} v_{\A} p_{s,\B} v_{\B} \\
          &- \phi \mu k \left( p_{s,\A} v_{\A} + p_{s,\B} v_{\B} \right) + \mu^2 = 0,
  \end{aligned}
\end{equation*}
yields, for each $s$, an equation containing the same two unknown variables: $v_{\A}$ and $v_{\B}$.
As the coefficients $p_{s,\A}$ and $p_{s,\B}$ will, in general, be different for each strain $s$,
one can see that, except for degenerate choices of the parameter values, no solution will exist for
more than two strains.

More specifically, we can see that, in the absence of specialists, small generalist populations can
grow in the two-patch approximation of this specific model if and only if $p_\xg \phi > \mu$, and
conversely that small populations of either specialist strain can grow in the absence of the
generalist if and only if $k \phi > \mu$.  Outside these regions, shown in figure
\ref{fig-meanfield}, the respective strains are not viable and will always die out (as the per
capita growth rates in this model are always maximized at vanishing population densities).

Within the region where all strains are viable, the approximated model has (in general) two possibly
stable equilibria: one where only the generalist is present, and one where the generalist is absent
and both specialists present.  (Any equilibria with only one specialist present are obviously
unstable against invasion by the other specialist.)  The former is stable if and only if $p_\xg >
k$, while the latter is stable if and only if $p_\xg < k$.  Only at exactly $p_\xg = k$, shown as
the white line in figure \ref{fig-meanfield}, both of the equilibria are neutrally stable, and are
connected by a line of neutrally stable equilibria along which neutral coexistence can occur.

\section{Simulation results}
Studying the dynamics of the full, unapproximated model requires numerical simulations.  As such
simulations tend to be computationally intensive, I have carried them out using custom, optimized
programs written in the C programming language.\footnote{Source code available from author under an
  open source license.}  The simulation code used for this paper includes two variants of the
coupling-based simulation algorithm described in \citep{couplingwip}, one using an occupancy list
for low population densities, and another using a vacancy list for high densities, with the outer
simulation loop periodically checking the population density and switching to the variant with the
higher mean time step per iteration.  I have also ported the basic simulation code (without the
coupling technique) to Java for demonstration purposes using interactive applets.

All simulation runs for this paper were done on a square $256 \times 256$ lattice with 8 neighbors
per site and with the edges wrapping around to the opposite sides.  In all simulation runs, time has
been scaled so that the \textit{per capita} mortality rate $\mu = 1$; in effect, I measure time in
mean individual lifetimes.  Habitat configurations were generated using an annealing method as
described in section \ref{landscape}.  The ``flea'' pseudorandom number generator \citep{flearng}
was used to produce random numbers, although I also carried out tests using other random number
generators to check that the results did not depend on such details.

\begin{figure}
  \centering
  \includegraphics[width=0.48\textwidth]{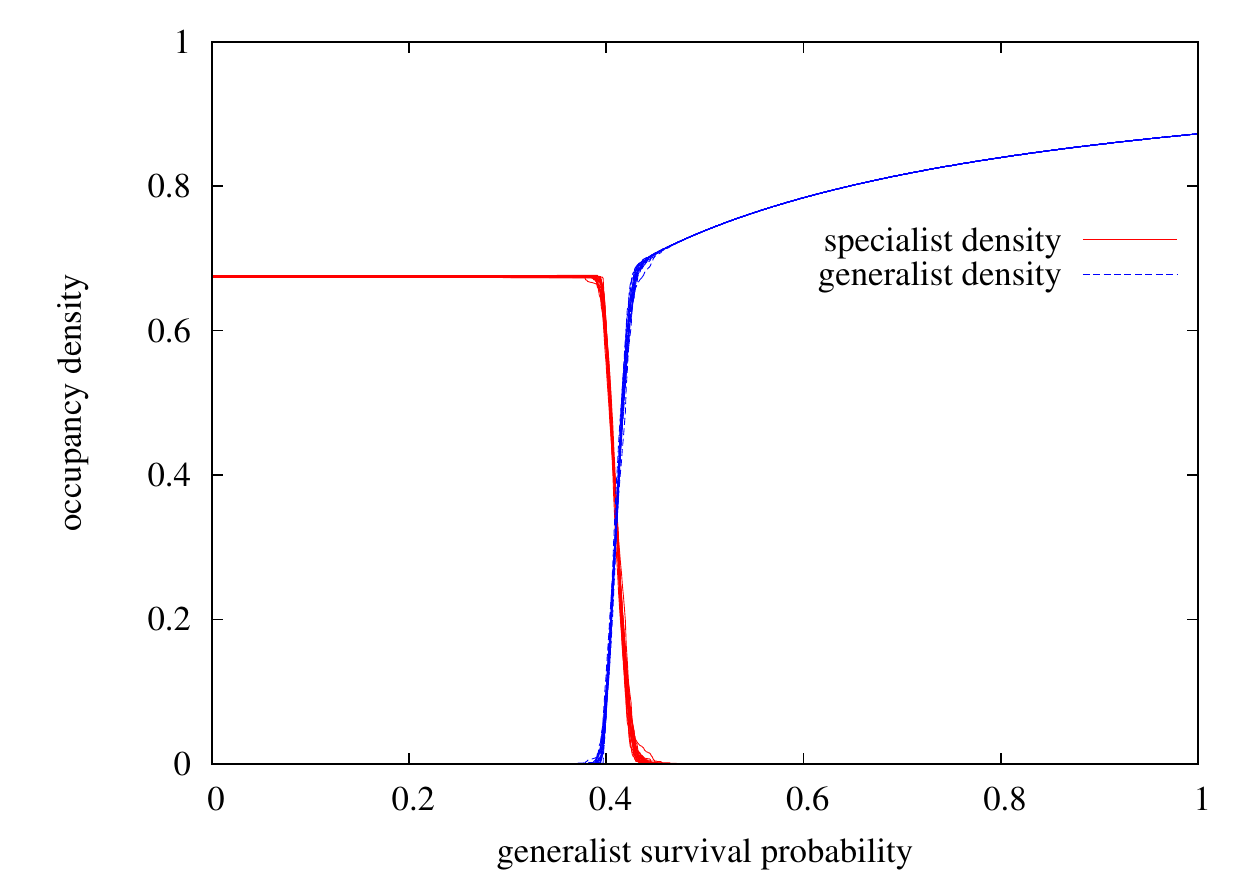}
  \includegraphics[width=0.48\textwidth]{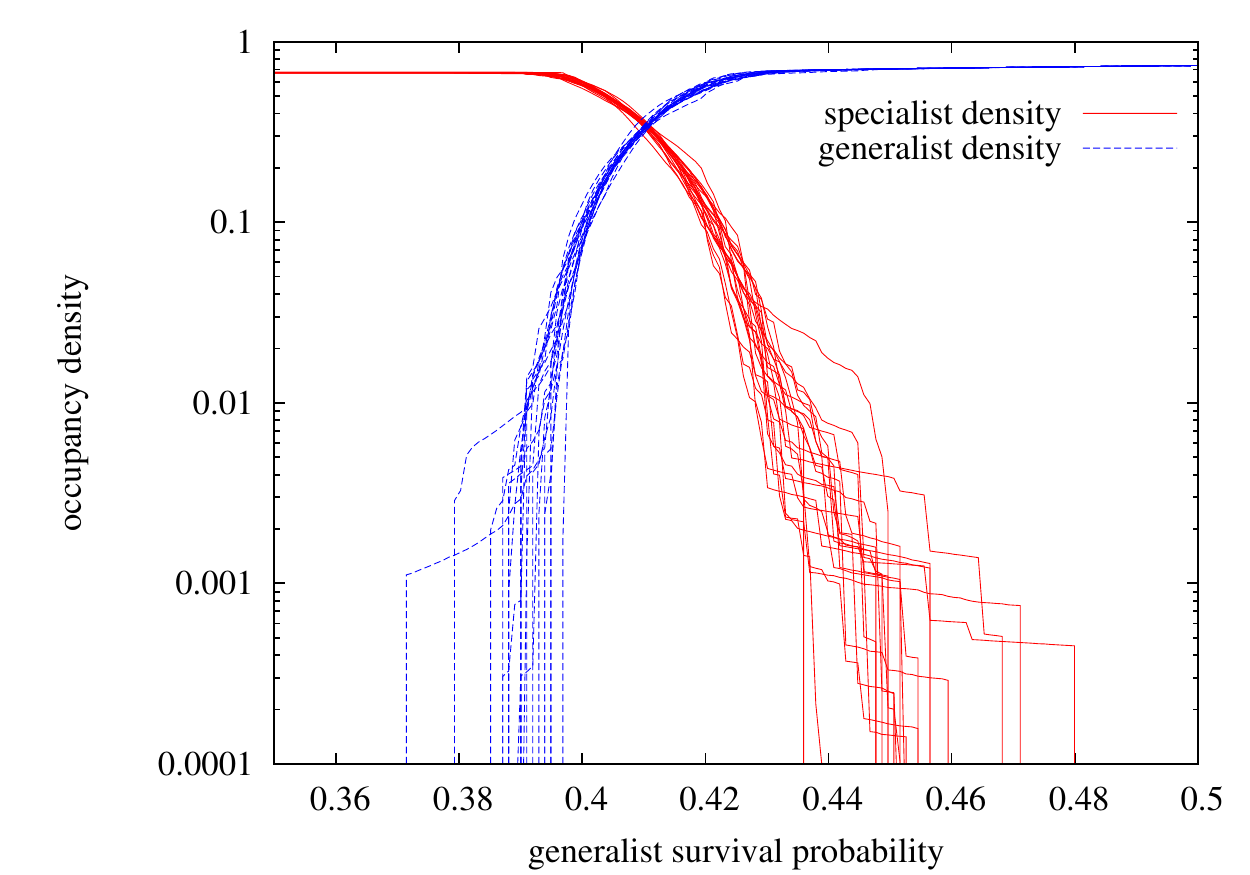}
  \caption{Plots of equilibrium specialist and generalist densities at $k=0.5$ and $\phi = 8\mu$ as
    functions of $p_\xg$ obtained from 20 numerical simulation runs.  See text for details.  The
    graph on the right has been plotted on a logarithmic scale and zoomed in to better show the
    coexistence region.}
  \label{fig1sim}
\end{figure}

Figure \ref{fig1sim} shows the equilibrium densities of generalists and specialists observed in
repeated individual-based simulations of the model on a $256 \times 256$ lattice with reproduction
to 8 nearest neighbors and wrapped edges with an uncorrelated habitat distribution and a moderate
value of $\phi/\mu$.  Each simulation run was started from a random habitat configuration and a
random initial state, with half the sites occupied by generalists and half by specialists.
Populations were allowed to equilibrate for $50000/\mu$ time units, after which population densities
were averaged over another $50000/\mu$ time units.  The specialist occupancy fractions are summed
over both specialist strains.

\begin{figure}
  \centering
  \includegraphics[width=0.45\textwidth]{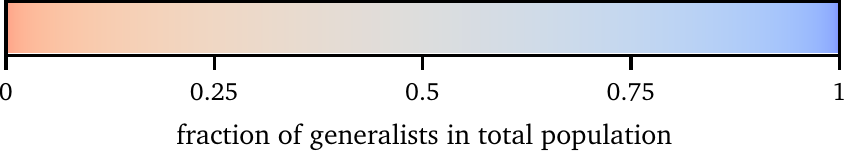} \\  
  \subfloat[$k=0.3$]{\label{fig2simA}
    \includegraphics[width=0.45\textwidth]{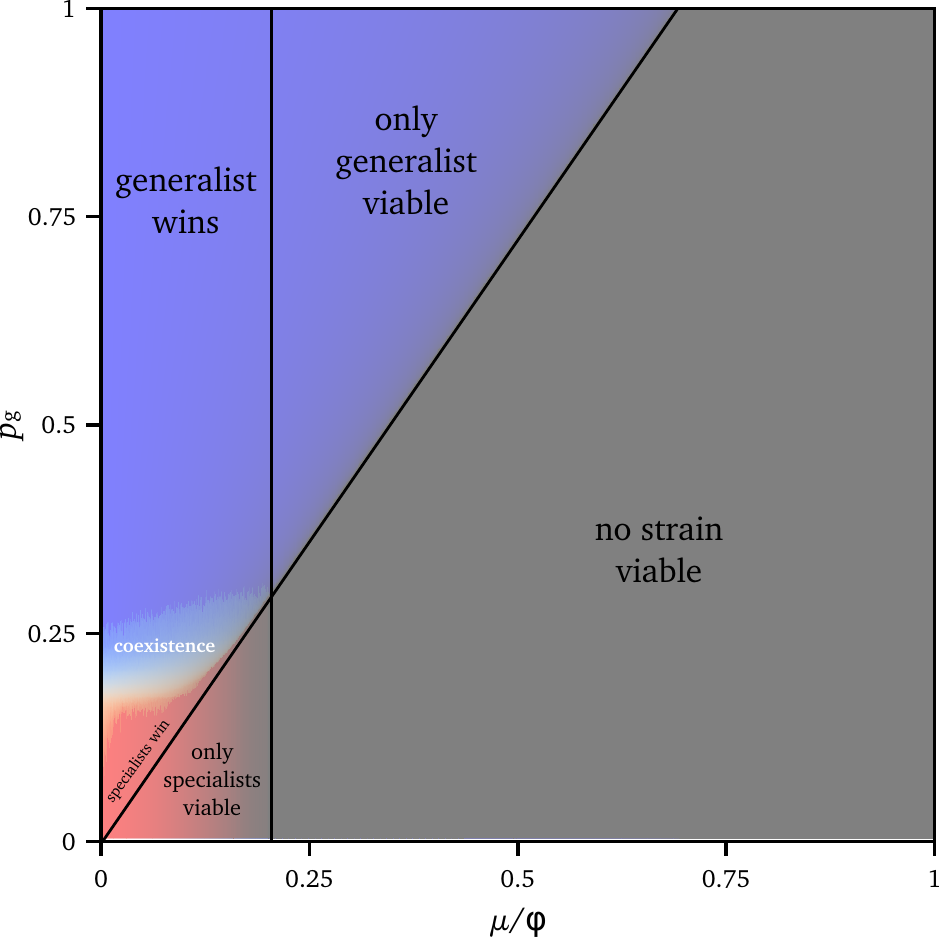} }
  \subfloat[$k=0.5$]{\label{fig2simB}
    \includegraphics[width=0.45\textwidth]{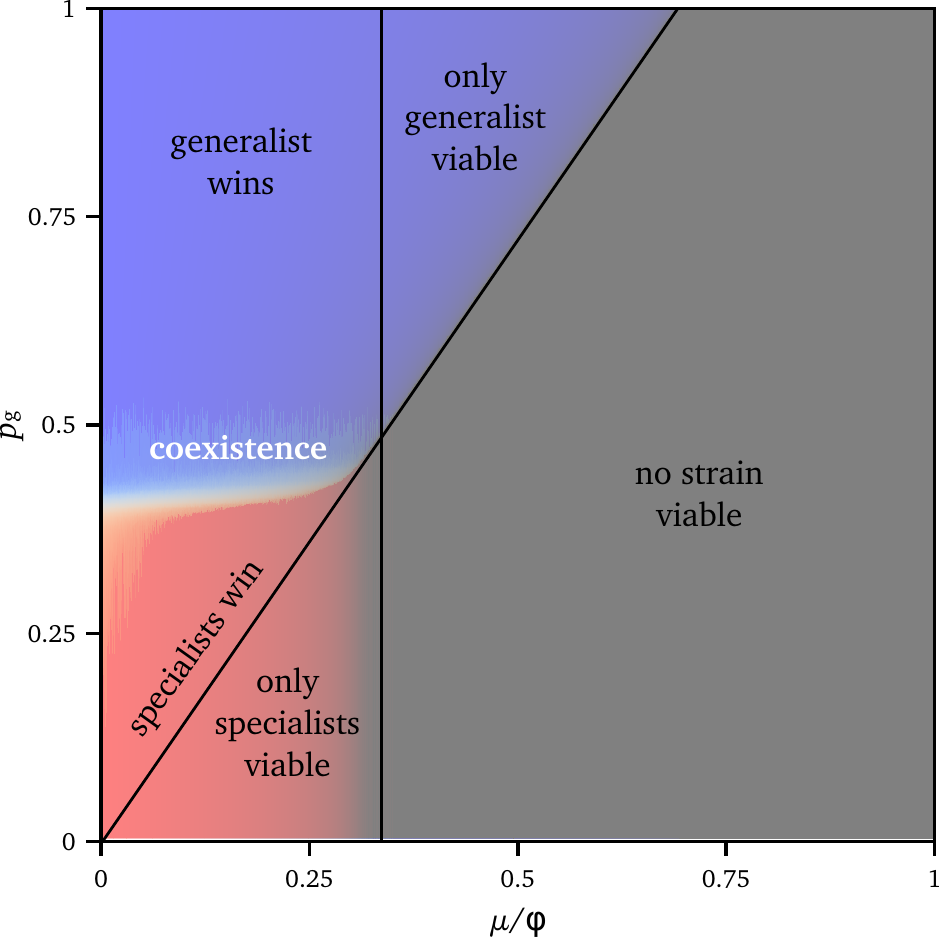} } \\
  \subfloat[$k=0.75$, $\gamma=10$]{\label{fig2simC}
    \includegraphics[width=0.45\textwidth]{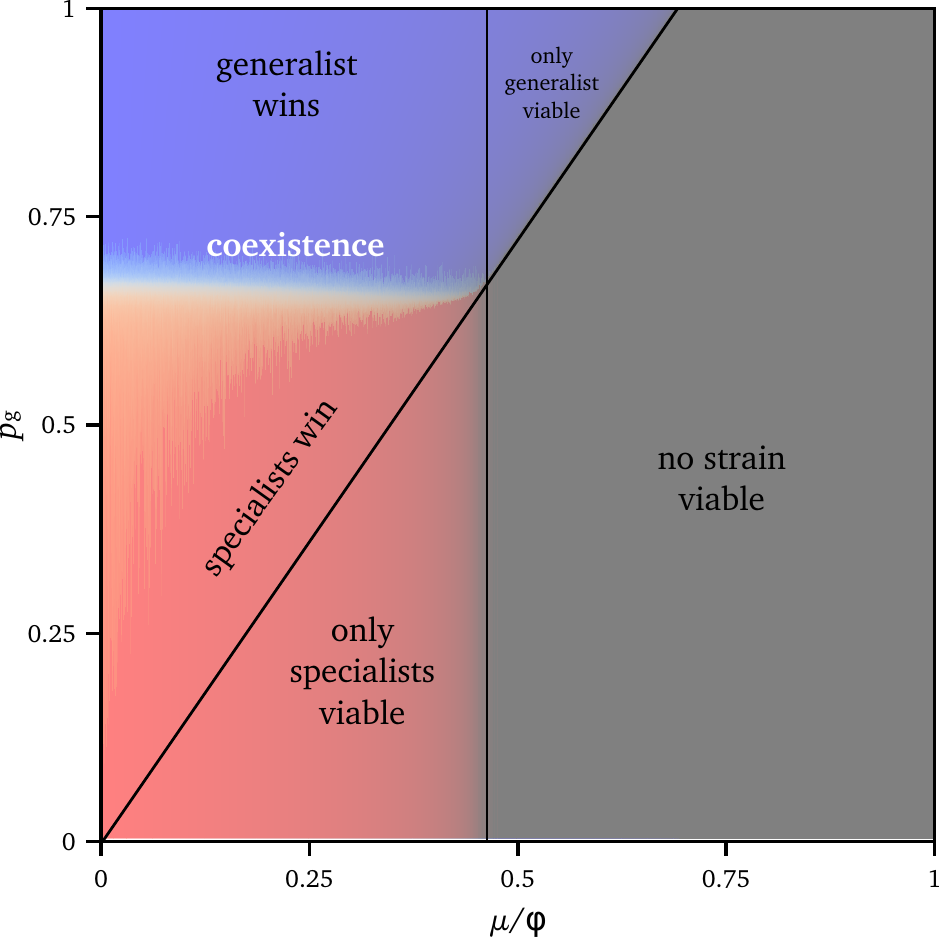} }
  \subfloat[$k=0.75$, $\gamma=3$]{\label{fig2simD}
    \includegraphics[width=0.45\textwidth]{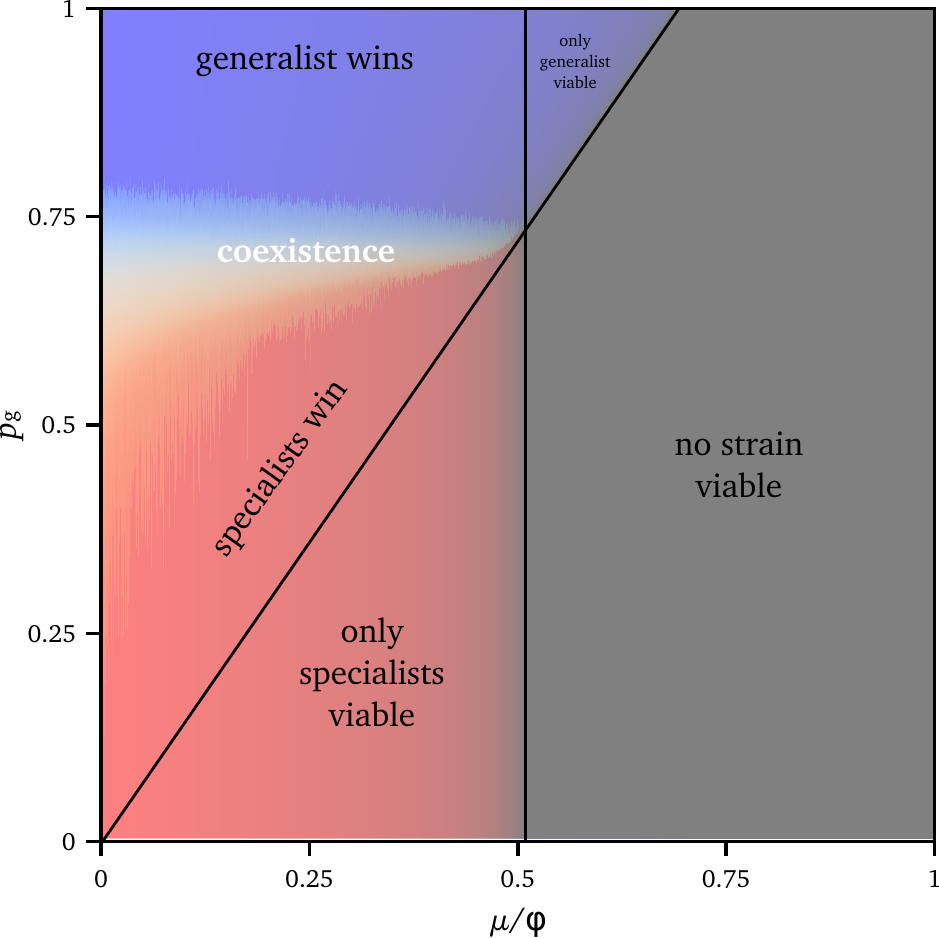} }
  \caption{Results of numerical simulations on a $256 \times 256$ site lattice as functions of
    $\mu/\phi$ and $p_\xg$ on the habitat patterns from figure \ref{fig3sim} with no global
    dispersal ($\epsilon = 0$).  See text for details.  Compare with figure \ref{fig2bsim} and with
    the mean field predictions from figure \ref{fig-meanfield}.}
  \label{fig2sim}
\end{figure}

\begin{figure}
  \centering
  \includegraphics[width=0.45\textwidth]{gradient.pdf} \\  
  \subfloat[$k=0.3$]{\label{fig2bsimA}
    \includegraphics[width=0.45\textwidth]{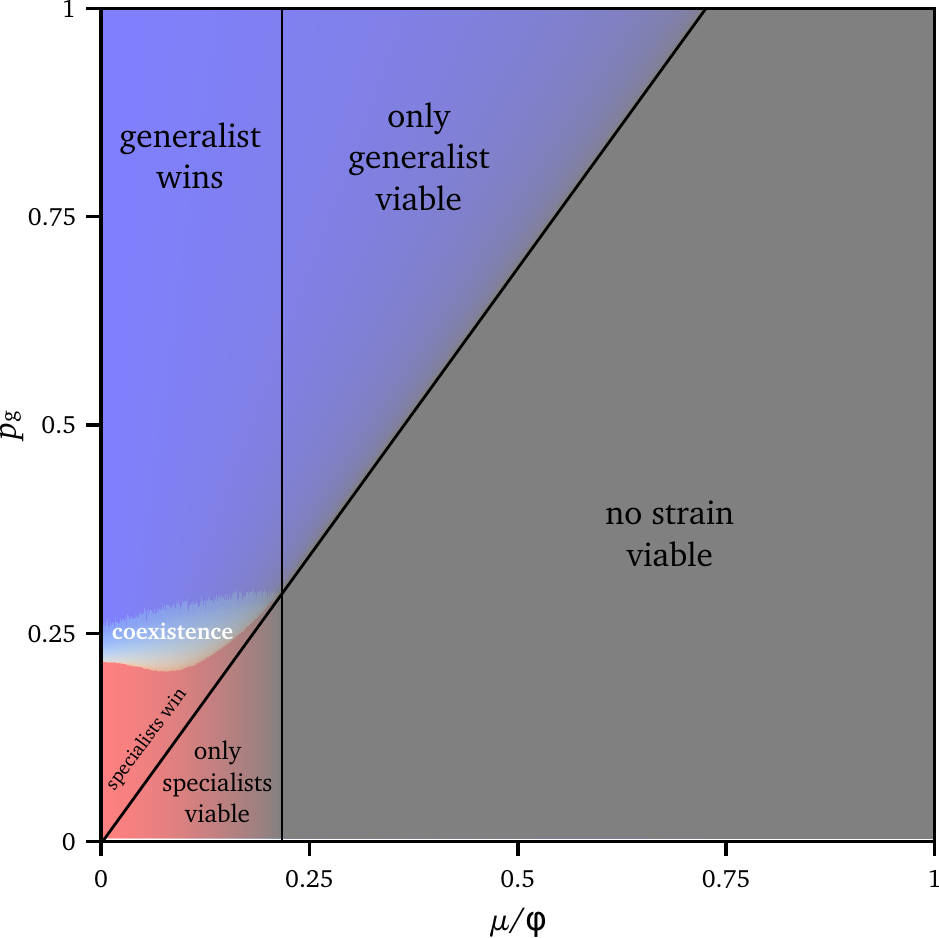} }
  \subfloat[$k=0.5$]{\label{fig2bsimB}
    \includegraphics[width=0.45\textwidth]{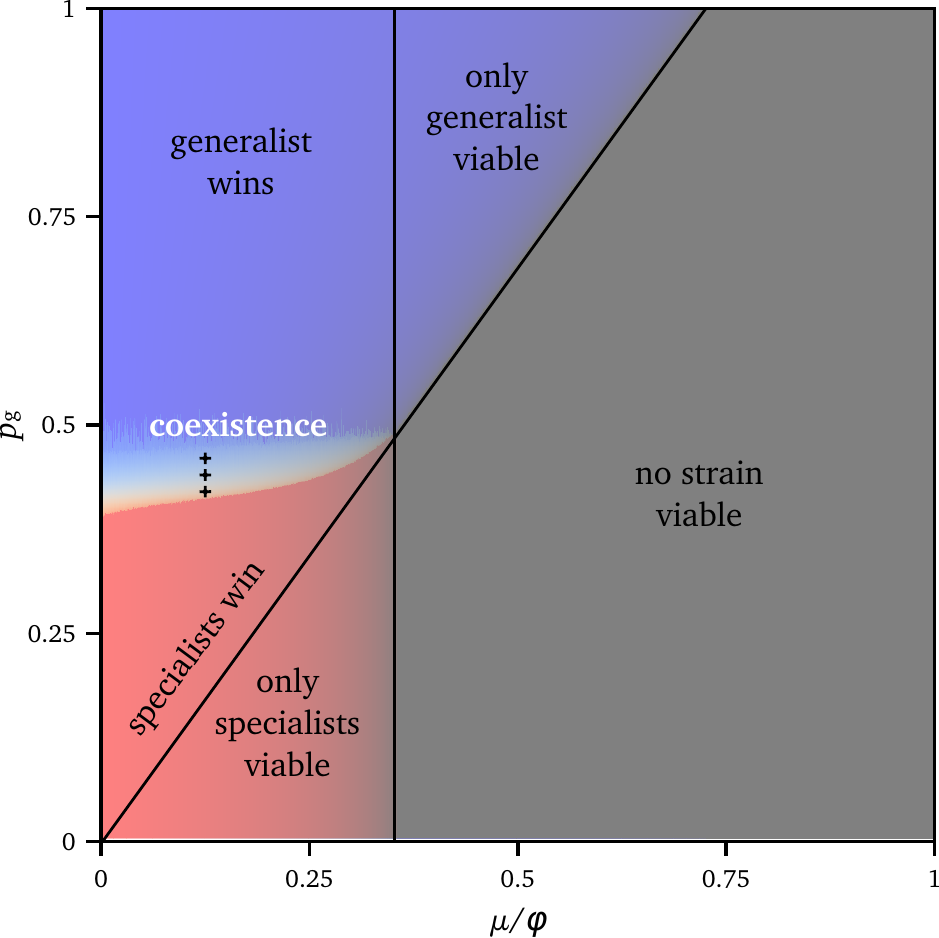} } \\
  \subfloat[$k=0.75$, $\gamma=10$]{\label{fig2bsimC}
    \includegraphics[width=0.45\textwidth]{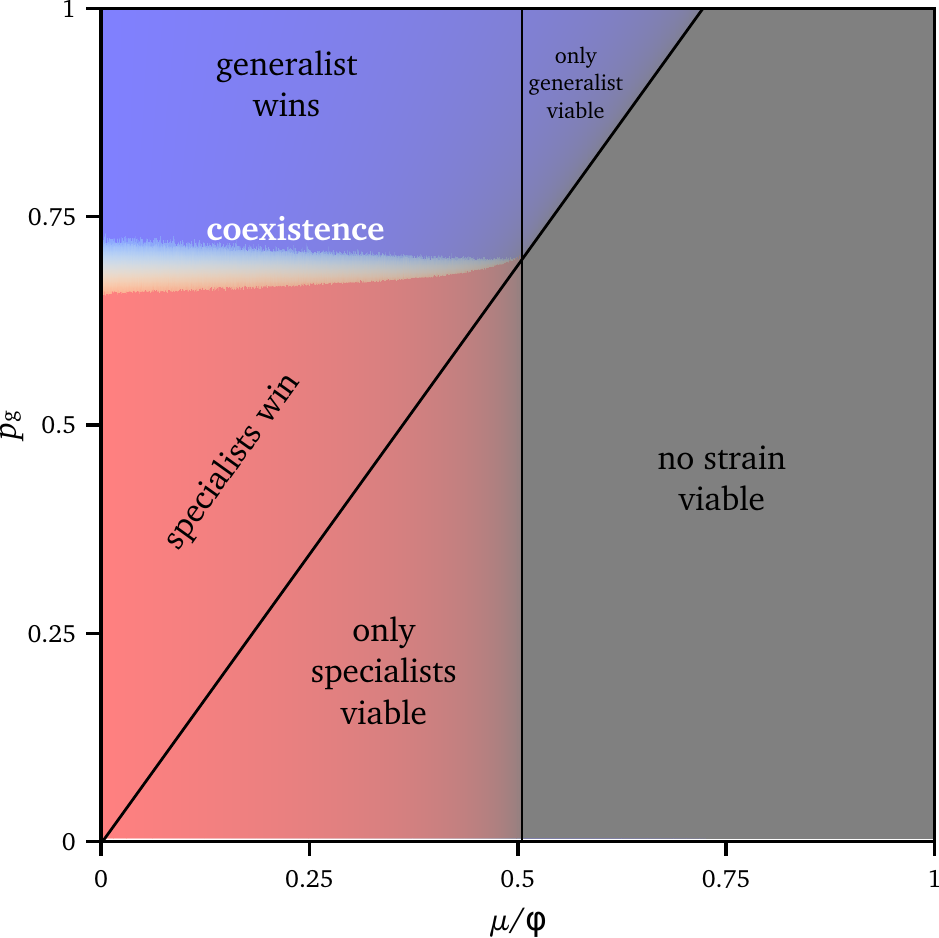} }
  \subfloat[$k=0.75$, $\gamma=3$]{\label{fig2bsimD}
    \includegraphics[width=0.45\textwidth]{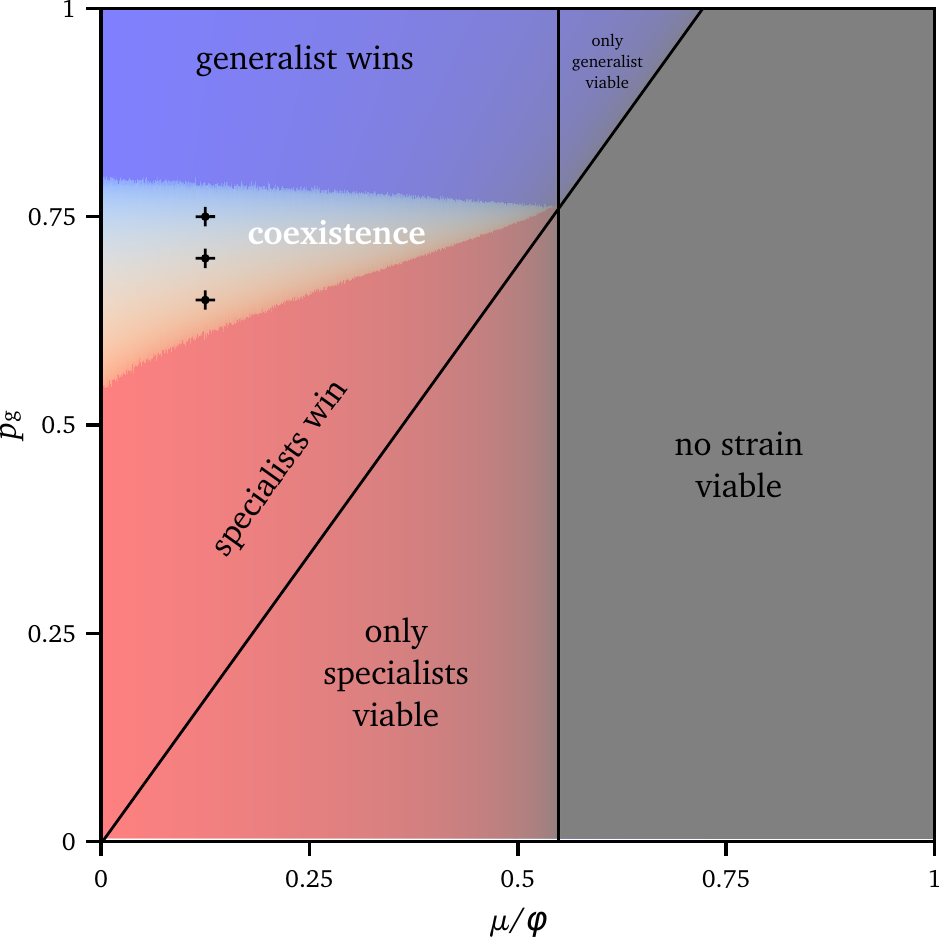} }
  \caption{Results of numerical simulations on a $256 \times 256$ site lattice as functions of
    $\mu/\phi$ and $p_\xg$ on the habitat patterns from figure \ref{fig3sim} with occasional global
    dispersal ($\epsilon = 0.01$).  See text for details.  The marks on figures \ref{fig2bsimB} and
    \ref{fig2bsimD} show the parameter values used for the invasion simulations in figure
    \ref{fig1inv}.  Compare with figure \ref{fig2sim} and with the mean field predictions from
    figure \ref{fig-meanfield}.}
  \label{fig2bsim}
\end{figure}

\begin{figure}
  \centering
  \subfloat[$k=0.3$]{\label{fig3simA}
    \includegraphics[width=0.45\textwidth]{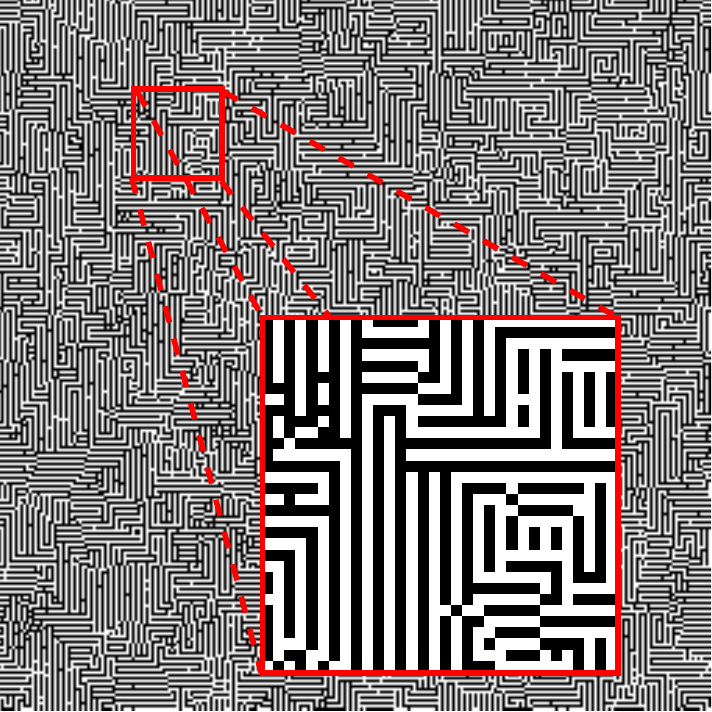} }
  \subfloat[$k=0.5$]{\label{fig3simB}
    \includegraphics[width=0.45\textwidth]{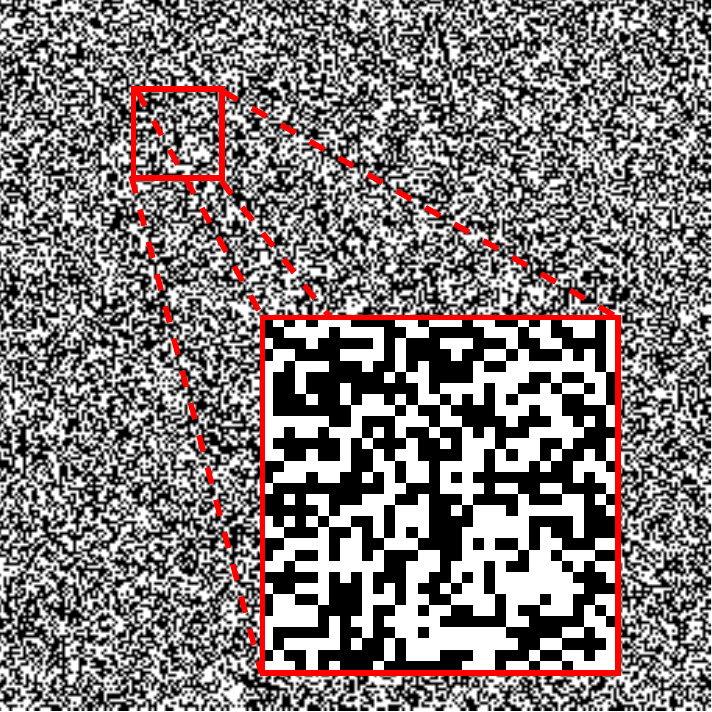} } \\
  \subfloat[$k=0.75$, $\gamma=10$]{\label{fig3simC}
    \includegraphics[width=0.45\textwidth]{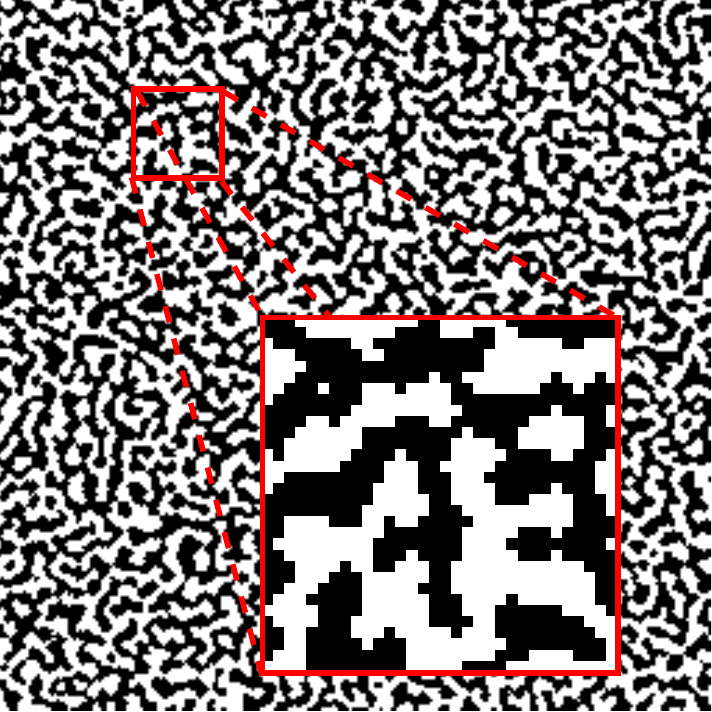} } 
  \subfloat[$k=0.75$, $\gamma=3$]{\label{fig3simD}
    \includegraphics[width=0.45\textwidth]{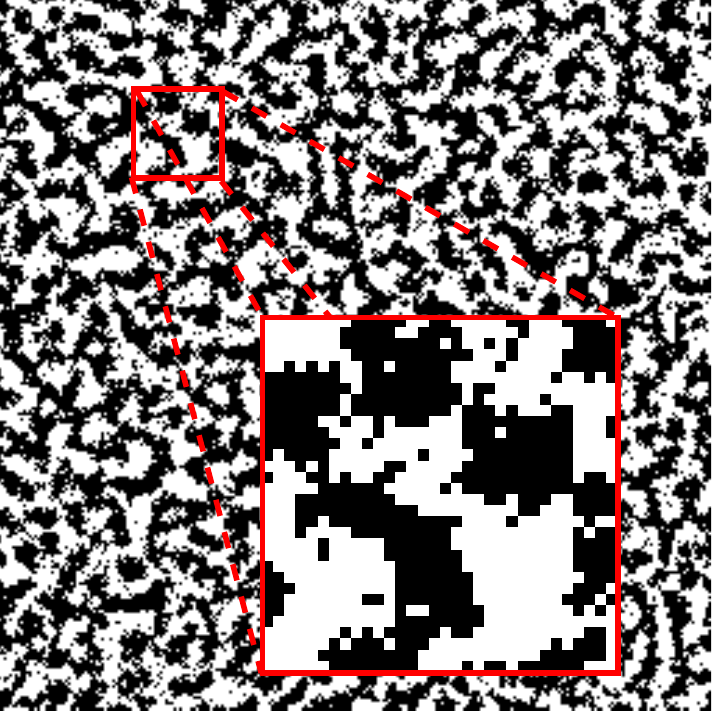} } 
  \caption{The habitat configurations used for the simulations shown in figures \ref{fig2sim} and
    \ref{fig2bsim}.  White squares correspond to habitat $\A$, black squares to habitat $\B$.  The
    insets show a $32 \times 32$ region magnified.  All lattices were generated from the same random
    initial state (which is nearly identical to lattice \ref{fig3simB}; only a very small amount of
    annealing was needed to make $k$ exactly $0.5$) using the annealing method described in section
    \ref{landscape}.  Lattices \ref{fig3simC} and \ref{fig3simD} have the same pairwise correlation
    number $k=0.75$, but the different annealing parameters used to generate them lead to visibly
    different higher order correlations and to corresponding differences in population dynamics.}
  \label{fig3sim}
\end{figure}

Contrary to the predictions from the mean field approximation, a non-degenerate region of the
parameter space where all three strains stably coexist can be seen in figure \ref{fig1sim}.  This
region is displayed more extensively in figures \ref{fig2sim} and \ref{fig2bsim}, which plot the
observed region of coexistence against $\mu/\phi$ and $p_\xg$ for the various habitat configurations
(anticorrelated, uncorrelated and two positively correlated patterns) shown in figure \ref{fig3sim}.
Figure \ref{fig2sim} shows results for pure nearest-neighbor dispersal ($\epsilon = 0$), while in
figure \ref{fig2bsim}, 1\% of all offspring were permitted to disperse uniformly over the whole
lattice ($\epsilon = 0.01$).

For figures \ref{fig2sim} and \ref{fig2bsim}, each simulation run was started from a random
initial population on the fixed pregenerated habitat landscapes shown in figure \ref{fig3sim}.
Populations were allowed to equilibrate for $20000/\mu$ time units, after which population densities
were averaged over $2000/\mu$ time units.  The red areas show where only the specialist strains
survived, while in the blue regions only the generalist strain remained.  The lighter shaded area
between them shows the part of the parameter space where both specialists and generalists survived,
with the color gradient shown above the figures indicating the relative average population
densities.

In both figures \ref{fig2sim} and \ref{fig2bsim}, the region of stable coexistence can be seen as a
more or less wedge-shaped area starting from the point where the viability boundaries of the strains
intersect, which is where the two-patch approximation would predict a line of neutral coexistence
(see figure \ref{fig-meanfield}).  The main difference between the figures can be seen in the lower
left side of the coexistence region: with global dispersal, the lower boundary of the coexistence
region is quite sharp, whereas with no global dispersal and high baseline fecundity $\phi/\mu$, the
generalist can often survive in small numbers (shown as a light orange hue in the plots) even where
the specialist dominates.

This happens simply because the high fecundity allows even small isolated population clusters to
survive for a long time, and because the strictly local dispersal prevents the specialists from
recolonizing isolated habitat patches.  If such a habitat patch happens to end up with no specialist
individuals (either because all happen to die out, or because none were present initially), the
patch can be colonized by generalists, which are then safe from competition there.  Allowing a
fraction of offspring to disperse globally lets the specialist strains recolonize such patches,
eliminating this effect.

It can also be seen that the addition of global dispersal generally reduces the width of the
coexistence region somewhat, although (except for the isolated patch effect noted above) the
reduction is not yet very large for $\epsilon = 0.01$.  This is to be expected, given that at
$\epsilon = 1$ the coexistence region reduces to a line, as shown by the mean field (and two-patch)
approximation above.

The results shown in figures \ref{fig1sim}, \ref{fig2sim} and \ref{fig2bsim} were calculated using a
simulation technique based on monotone coupling \citep{couplingwip}, which allows the system to be
efficiently simulated for all values of the parameter $p_\xg$ in parallel.  Each line in figure
\ref{fig1sim} and each vertical stripe (out of 1024 per plot) in figures \ref{fig2sim} and
\ref{fig2bsim} corresponds to one simulation run.  Because the simulation technique used causes the
effects of random demographic fluctuations on populations with different values of $p_\xg$ within
the same run to be correlated, the results show stronger correlations within each run than between
runs, which should be kept in mind when interpreting the figures.

\begin{figure}
  \centering
  \subfloat[$k=0.5$, $p_\xg = 0.43$]{\label{fig4simA}
    \includegraphics[width=0.3\textwidth]{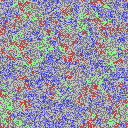} }
  \subfloat[$k=0.75$, $p_\xg = 0.7$]{\label{fig4simB}
    \includegraphics[width=0.3\textwidth]{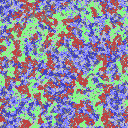} } 
  \subfloat[$k=0.95$, $p_\xg = 0.93$]{\label{fig4simC}
    \includegraphics[width=0.3\textwidth]{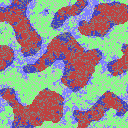} }
  \caption{Snapshots of the population equilibrium after a few hundred mean lifetimes for various
    values of $k$, with $p_\xg$ near the middle of the coexistence region.  Snapshots were taken
    from simulations run on a $128 \times 128$ lattice with $\phi = 4\mu$ and $\epsilon = 0.001$.
    The red and green sites are occupied by $\xa$ and $\xb$ specialists respectively, the blue sites
    are occupied by generalists and the gray sites are vacant.  For the blue and gray colors, darker
    shades are used for habitat A and lighter shades for habitat B.  These snapshots have been taken
    from interactive Java applets available at \url{http://vyznev.net/ca/coex2env/}.}
  \label{fig4sim}
\end{figure}
Figure \ref{fig4sim} contain snapshots of simulations run on lattices with different site type
patterns.  It can be seen that, when sites of the same type are strongly clustered, large
contiguous clusters are dominated by the respective specialist strain, while the generalist strain
is able to persist in areas near cluster boundaries and in isolated minor clusters too small to
support a stable specialist population.

On the other hand, when site types are uncorrelated, a different pattern is observed.  Such
lattices contain no large contiguous clusters that could be dominated by one specialist strain;
instead, the two specialist strains tend to occur together in regions where the random distribution
of site types happens to favor one or both of them.  Through competition with the generalist
strain, the two specialist strains indirectly support one another, even though there is no direct
interaction between them.

\section{Mutual invasibility}

To a skeptical mind, the results presented above may yet leave some doubt about whether the apparent
coexistence observed in these simulations is indeed genuinely stable, or merely an artifact of slow
convergence and insufficient simulation time.  After all, if the simulation was run for long enough
on a finite lattice, \emph{eventually} one of the strains (and eventually all of them) would almost
surely go extinct simply due to demographic stochasticity.  Thus, it may not even be entirely clear
what ``stable coexistence on a finite lattice'' should actually mean.

On an infinite lattice, as in \citet{lanchier2006}, a set of strains may be said to coexist stably
if they can all persist indefinitely long with non-zero probability.  By this definition, no stable
coexistence (or even just existence) is possible in a finite system.  However, since there do exist
known results that relate the scaling of the expected time to extinction on a finite lattice, as a
function of lattice size, to the limiting behavior of the model on an infinite lattice\TODO{ref
  liggett}, one might be inclined to try and use such scaling laws to extrapolate stability from
small lattices to the infinite limit.

This is not the approach I have taken.  After all, real habitats and populations, like the
simulations employed in this paper, are finite --- in appealing to a definition of coexistence that
only works for infinite populations, one ends up obscuring the fact that, in reality, if a
population of tens of thousands of individuals persisting over equally many generations is not
considered stable, it's hard to say what should be.

Rather, I wish to demonstrate the stability of the trimorphic coexistence in my model in a more
direct manner, by showing that it exhibits \emph{mutual invasibility}.  That is to say, if a small
number of individuals of any of the three strains are introduced into a stable population consisting
solely of the other two strains, the introduced strain will, with positive probability, survive and
grow in number up to its equilibrium density in the trimorphic equilibrium, with the initial part of
the growth curve appearing approximately exponential.

\begin{figure}
  \centering
  \subfloat[generalist invasion ($k=0.5$)]{ \label{fig1invA}
    \includegraphics[width=0.49\textwidth]{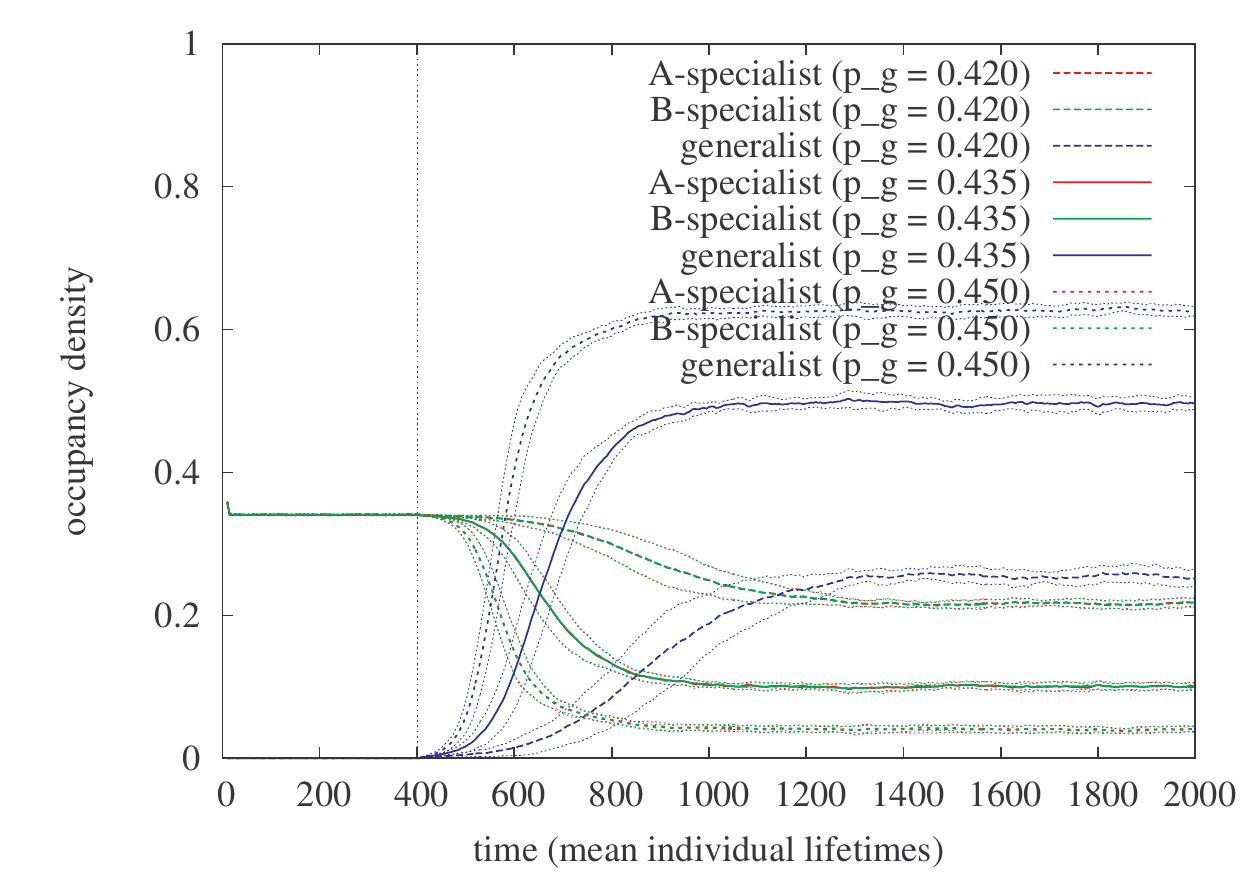} } 
  \subfloat[specialist invasion ($k=0.5$)]{ \label{fig1invB}
    \includegraphics[width=0.49\textwidth]{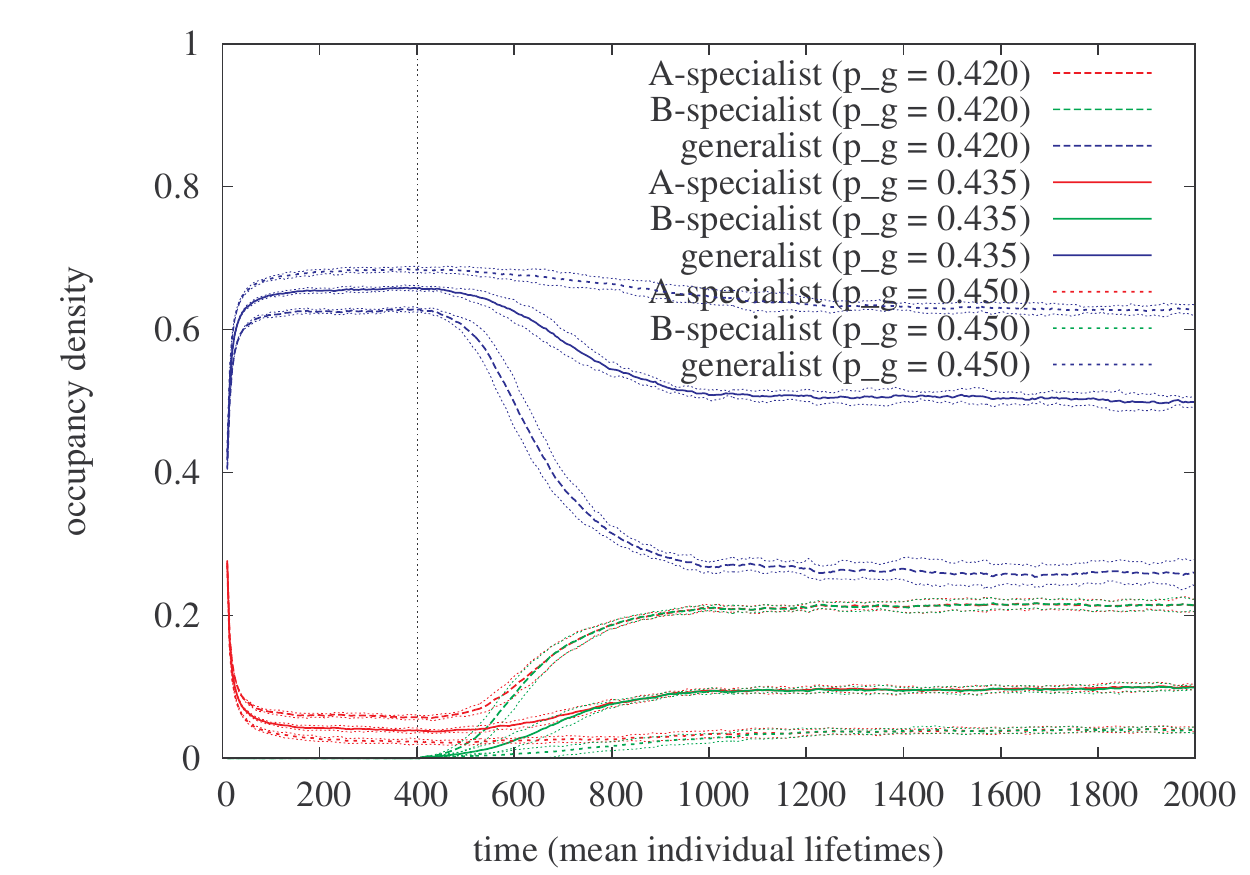} } \\
  \subfloat[generalist invasion ($k=0.75, \gamma=3$)]{ \label{fig1invC}
    \includegraphics[width=0.49\textwidth]{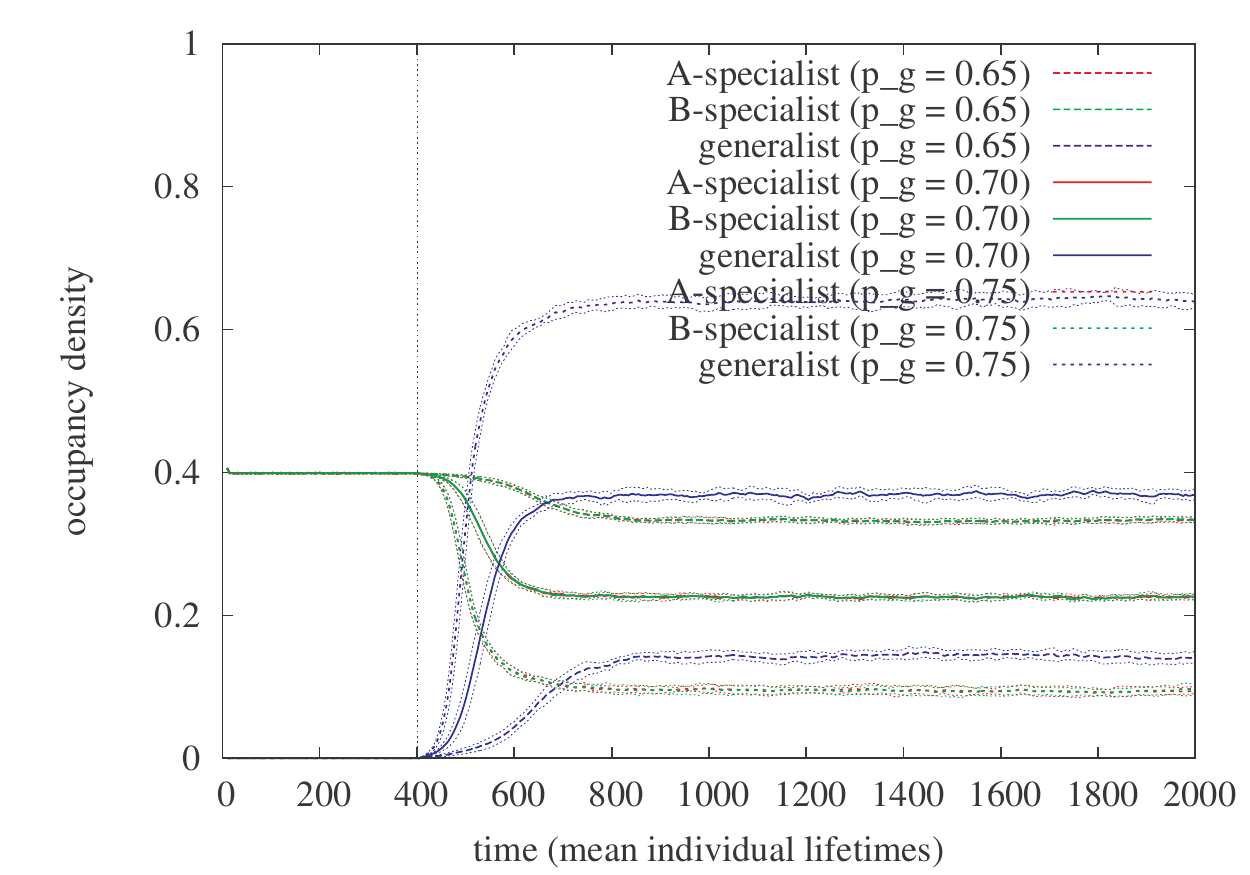} } 
  \subfloat[specialist invasion ($k=0.75, \gamma=3$)]{ \label{fig1invD}
    \includegraphics[width=0.49\textwidth]{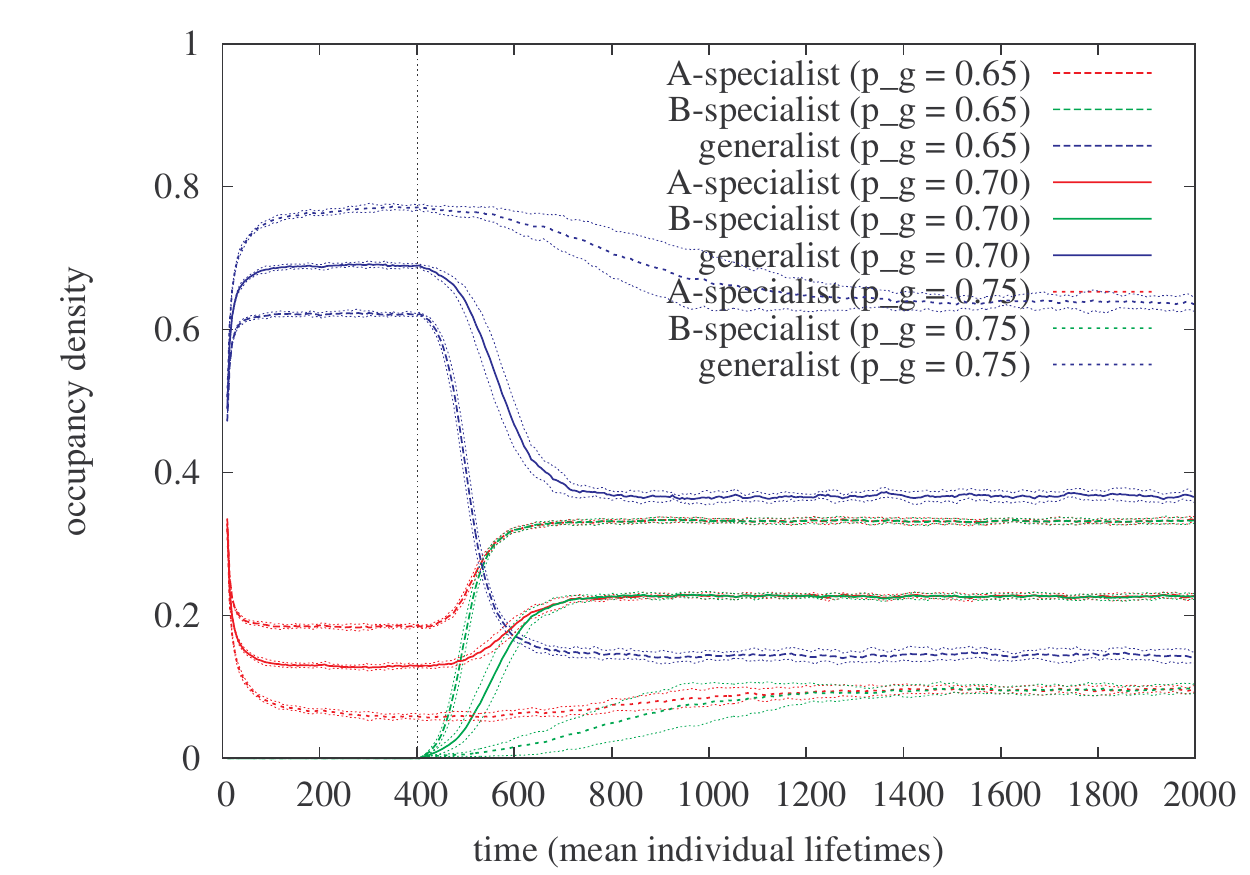} } \\
  \caption{Simulations showing mutual invasibility at the parameter values marked in figures
    \ref{fig2bsimB} and \ref{fig2bsimD}; $\phi = 8 \mu$ for all simulations.  A dimorphic population
    is allowed to equilibrate for $400/\mu$ time units, after which $100$ individuals (i.e. initial
    density $\approx 0.0015$) of the third strain are placed randomly on the lattice.  In
    simulations of specialist invasion, the invading strain is without loss of generality taken to
    be $\xb$.  The population densities shown in the graphs are smoothed over $5/\mu$ time units and
    averaged over $10$ independent simulation runs; invasion does succeed in all runs.  The red,
    green and blue lines show the average $\xa$, $\xb$ and $\xg$ strain densities respectively,
    while the solid, dashed and dotted lines correspond to different values of $p_\xg$ within the
    coexistence region.  (In populations with both specialist strains present, the red and green
    lines tend to overlap.)  The thin dotted lines are drawn one sample standard deviation above and
    below the corresponding average line.}
  \label{fig1inv}
\end{figure}

Figure \ref{fig1inv} shows some simulations demonstrating mutual invasibility at six points within
the coexistence region of the parameter space.  With 100 initial invader individuals, the invader
strain survived and established itself in all runs carried out --- evidently the invasion
probability of a single invader exceeds $1/100$ at all the sampled parameter values.  The population
density of the invading strain over time shows a distinctive sigmoid shape, with the initial growth
being approximately exponential.  A notable feature visible in the plots is that invasion by a
specialist strain also increases the population of the other specialist strain already present; this
happens because both specialists are in competition with the generalist strain.  The relatively high
variance seen in some of the plots during the growth phase is due to initial demographic
stochasticity affecting the time until exponential growth sets in; once properly started, the shape
of the growth curve is very similar in all runs.

Were the coexistence observed in this model merely neutral, the population density of a newly
introduced strain would be as likely to decrease as to increase as the result of stochastic
fluctuations.  The fact that, at the sampled parameter values, small populations of each strain
instead show a clear increasing trend confirms that this model exhibits true, non-neutral
coexistence.

\section{Discussion}
In this paper I've demonstrated, using a simple toy model of competitive population dynamics on a
lattice, that spatial heterogeneity is one of the mechanisms by which the competitive exclusion
principle can be violated.  The fact that this cannot occur in well-mixed populations shows that
population viscosity and explicit spatial structure are essential to this mechanism.

Had the model included more than two habitat types, temporal variation, hierarchical competition or
nonlinear interactions between individuals, the coexistence of multiple strains would not have been
at all surprising.  Yet it has none of these, and can still support more than two strains in stable
coexistence.  All that allows such coexistence to persist in this model is the combination of
environmental variation, persistent spatial structure and distance-limited dispersal; eliminating
any of these reduces the model to one capable of supporting no more strains than would be predicted
by a naive application of the competitive exclusion principle.

Real organisms do not usually live in a completely homogeneous environment, nor do most of them
disperse uniformly over their entire habitat.  It is obvious and commonly acknowledged that
environmental variation can increase diversity, yet the fact that, when combined with
distance-limited dispersal, this increase can be more than linear seems to have attracted little
attention.  Yet the ubiquity of habitat edges and fragmented landscapes in nature suggests that it
should be possible to find examples of this type of coexistence in nature, and indeed that such
``edge effects'' may contribute to the generation and maintenance of ecological diversity in many,
if not most, ecosystems.

I find, however, that in many ways this work has raised more questions than it has answered.  For
example, an obvious question would be whether the model allows the stable coexistence of more than
three strains.  Another natural question is whether the coexistence of three or more strains in this
type of model can also be evolutionarily stable, and further, whether it might arise from a mono- or
dimorphic state through evolutionary branching \citep{geritz1998,magori2005}.  Based on limited
simulation experiments, the answer to all of these questions appears to be ``yes'', although the
conditions still need to be explored more thoroughly.

\section{Acknowledgments}

I would like to thank my colleague Robert Service for his comment during a presentation which
originally led me to investigate this model, and my advisor Éva Kisdi for her guidance and
assistance.  I am also grateful to Minus van Baalen for useful discussions and for suggestions
regarding the landscape generation algorithm, and to the editor and reviewers of the previous
revisions of this paper for their valuable feedback.

This work was financially supported by the Finnish Graduate School in Computational Sciences
(FICS).

\bibliographystyle{model2-names}
\bibliography{coex2env}

\end{document}